\documentclass{cs23proc}

\usepackage{kantlipsum}
\usepackage{url}
\editors{Takeru Suzuki and the Cool Stars 23 Organizing Team}
\publisher{Zenodo}
\conference{The 23rd Cambridge Workshop on Cool Stars, Stellar Systems, and the Sun (Cool Stars 23)}
\conferencedate{2026}

\title{Multiplicity of Cool Stars and their Evolution}
\author{Matthew I. Swayne$^{1}$,
        Leen Decin$^{2}$,
        Theo Khouri$^{3}$,
        Ayush Moharana$^{4}$,
        Léa Planquart$^{3}$,
        John Southworth$^{4}$,
        Nikki J. Miller$^{5}$,
        Jos\'e A. Caballero$^{6}$,
        Orsolo De Marco$^{7,8}$,
        Krzysztof G. He{\l}miniak$^{9}$,
        Swetlana Hubrig$^{10}$,
        Valentin D. Ivanov$^{11}$,
        Pierre Kervella$^{12,13}$,
        Thibault Merle$^{14,15}$,
        Dorota M. Skowron$^{16}$}

\affiliation{$^{1}$ School of Physics \& Astronomy, University of Glasgow, United Kingdom \\
			 $^{2}$ Department of Physics and Astronomy, KU Leuven, Leuven, Belgium \\
			  $^{3}$  Department of Physics and Astronomy, Chalmers University of Technology, 41296 Gothenburg, Sweden\\
            $^{4}$ Astrophysics Group, Keele University, ST5 5BG, Staffordshire, UK\\
            $^{5}$ Department of Physics and Astronomy, Uppsala University, Box 516, 75120 Uppsala, Sweden\\
            $^{6}$ Centro de Astrobiología (CSIC-INTA), 
            Villanueva de la Cañada, Madrid, Spain\\
            $^{7}$ Department of Physics and Astronomy, Macquarie University, Sydney NSW 2109, Australia \\
            $^{8}$ Astrophysics and Space Technologies Research Centre, Macquarie University, Sydney NSW 2109, Australia\\
            $^{9}$  Nicolaus Copernicus Astronomical Center, Polish Academy of Sciences, ul. Rabiańska 8, 87-100 Toruń, Poland\\
            $^{10}$ Leibniz-Institut für Astrophysik Potsdam (AIP), An der Sternwarte 16, 14482 Potsdam, Germany\\
            $^{11}$ European Southern Observatory, Karl-Schwarzschild-Str. 2, 85748 Garching-bei-M\"unchen, Germany\\
            $^{12}$ LIRA, Observatoire de Paris, Université PSL, Sorbonne Univ., Univ. Paris Cité, CY Cergy Paris Univ., CNRS, 92190 Meudon, France\\
            $^{13}$ French-Chilean Laboratory for Astronomy, IRL 3386, CNRS and U. de Chile, Casilla 36-D, Santiago, Chile\\
            $^{14}$ Royal Observatory of Belgium, Avenue Circulaire 3, 1180 Brussels, Belgium\\
            $^{15}$ Institut d’Astronomie et d’Astrophysique, Université Libre de Bruxelles, CP 226 Boulevard du Triomphe, 1050 Bruxelles, Belgium\\
            $^{16}$ Astronomical Observatory, University of Warsaw, Al. Ujazdowskie 4, 00-478 Warszawa, Poland}

\shorttitle{Multiplicity of Cool Stars and their Evolution}
\shortauthors{Swayne et al.}

\abs{Making up a sizeable portion of the galactic census, stellar multiples are experiencing a renaissance.
Enabling the study of multiple strands of the study of cool stars, stellar multiples have been found and characterised in great numbers by the missions of the last decade, allowing the exploration of stellar parameters and populations, observation of stellar interactions, studies into stellar formation and evolution, and characterisation of circumbinary systems.
This exciting explosion of science is only set to continue, with future missions set to offer even further insights into the topic.
Within these proceedings we will summarise the presentations and discussions on cool stellar multiplicity within our splinter sessions at the 23rd Cambridge Workshop on Cool Stars, Stellar Systems, and the Sun, as we examine the present state of the field and look to what the future may bring.}

\begin{document}

\maketitle

\section{Introduction}
More than 20 per cent of all stars more massive than 0.1\,${\rm M_\odot}$ exist in a multiple system, with the fraction increasing with mass to around 50 per cent for main sequence solar-type stars \citep{DucheneKraus2013,Offner2023}. Therefore, to completely understand star-formation and stellar evolution, the study of stellar multiplicity is essential. Multiplicity studies have advanced significantly since the early years of binary star research \citep{1654Hodierna,1778Mayer,Herschel_1786}. Starting from identifying single systems, and crude descriptions of small samples, we have moved to detailed studies of multiplicity statistics and its dependency on mass, metallicity and its environment \citep{Raghavan_2010}.

Multiplicity of cool stars has several implications on both stellar and planetary evolution. Formation of binaries from the protostellar core affects the material available for subsequent planet formation and where it is located within the system \citep{Cuello_2025}. Upon formation, dynamical effects of multiple systems drive the orientation of the planetary orbits and also decide their orbital resonances \citep{Baycroft_2025}. Close binary stars are known to affect each other via various interactive phenomena such as mass transfer and tidal synchronisation \citep{Paczynski_1971}. These effects change the configuration of the stars over the course of their evolution and result in various evolutionary outputs that dramatically differ from single star evolution theory. However, if the stars in the binary are sufficiently separated to avoid interaction but also show eclipses, we can use them to extract precise direct measurements of stellar properties and use them to calibrate models for single star evolution especially if the age is known, for example through cluster membership.

 We are, at present, in a renaissance of multiplicity. This is visible in the exponential increase in the number of multiple systems registered in the 9th Catalogue of Spectroscopic Binary Orbits (SB9, \citealt{Pourbaix_2004}), the Multiple Star Catalog (MSC; \citealt{MSCToko1997, Tokovinin_2018}) and the Washington Double Star Catalog (WDS; \citealt{2001Mason}). 
SB9 is an historical catalogue of spectroscopic binaries (SB) that has grown from 2400 SB in 2004 to more than 4000 SB 20 years later.
MSC is a collection of stellar systems with more than two components. The catalogue listed 3000 systems in 2010 in comparison to 14,600 in its last 2024 update.
WDS is a database of visual and astrometric double and multiple star information. Containing over 150,000 systems, the WDS designation system was of insufficient precision for the number of systems created in the advent of large sky surveys.
This resulted in the creation of the Washington Double Star Supplemental Catalog (WDSS) which has grown to over 2,400,000 systems in the latest 2024 update \citep{2025Mason}.

One of the reasons for this increase is the exploitation of several large space-based missions and ground-based surveys. The largest number of detections have been through astrometry, especially {\it Gaia} \citep{2021ElB,Gaia_2023,Czavalinga_2023}. But the most prolific, in terms of detection and characterisation, are the photometric missions: {\it Kepler} \citep{Prsa_2011,2016MNRAS.455.4136B,Helminiak_2019} and {\it TESS} \citep{Prsa_2022,Kostov_2022,Mitnyan_2024}, which have allowed us to detect and characterise at least 10,000 new binaries \citep{2025Kostov}. Multiplicity-focused projects have additionally benefitted from large spectroscopic surveys such as from {\it Gaia}-ESO \citep{2012Gilmore,Merle2017}, APOGEE \citep{Kounkel_2021} and LAMOST \citep{Jing_2025} as well as photometric surveys such as OGLE \citep{2015Udalski}. This only looks to increase with multiple ground-based spectroscopic facilities in the works such as 4MOST \citep{2019Msngr.175....3D}, MUST \citep{2024arXiv241107970Z} and the WST \citep{2024arXiv240305398M} among others, whose time-domain observing strategies will allow a significant impact on the observational constraints of stellar multiplicity. Single-star evolution and population synthesis models have been moving towards inclusion of binary or higher multiplicity modules \citep{Paxton_2015,Toonen_2016,Hamers_2021}. 
Finally, the {\it JWST} has the potential to provide insights into the faintest ultracool dwarf binaries \citep{Marley2009}, having already begun to discover, resolve and characterise some of the coolest binaries to date \citep{2023Cal,2025Gagliuffi}.

This wealth of recent high-quality multiplicity data is of particular interest for the low-mass stars and substellar community.
It can allow us to better explore fundamental questions in the make-up and processes of these fainter objects, such as between different classes of brown dwarfs and the transition from partly to fully convective stars \citep{2017Smart,2023MaxtedEBLM}. 
With new cutting-edge missions and projects proposed and being developed for this decade and the next such as \textsl{PLATO} (PLAnetary Transits and Oscillations of stars, \citealt{Rauer2025}), Vera C. Rubin Large Synoptic Survey Telescope (Rubin LSST, \citealt{Hambleton_2023}) and the {\it Nancy Grace Roman Space Telescope} ({\it Roman}, \citealt{Fatheddin_2023}); the contribution of stellar multiplicity to the exploration of Cool Stars will only increase.

This exciting topic was explored in a splinter session at the Cool Stars 23 conference in Tokyo, Japan titled `Multiplicity in cool stars and their evolution'.
The splinter session consisted of two Blocks, one focusing on the current development and challenges within the field while the other looked to preparations and plans for future studies of cool stellar multiplicity.
In total incorporating five plenary talks, 11 contributed talks and a poster-pop session, the splinter provided a wide-ranging look at the topic.
In questions and an end-of-splinter discussion session, there was a lively debate and discussion on presenter's research and on the future direction of cool stellar multiplicity studies.
Slides from presenters can be found at Zenodo\footnote{\url{https://zenodo.org/communities/coolstars23/}} and individual proceedings may be found there as well as in this summary of the splinter.

In the following sections we will summarise the research presented at the splinter session, split into their respective Blocks by presentation type.
Section~2 will summarise the research presented in Block 1 `Multiplicity of cool stars -- Current Developments and Challenges', including three plenary talks, seven contributed talks and the `poster-pop' session.
Section~3 will summarise the research presented in Block 2 `Preparing for the future -- Future missions and Cool Stars Multiplicity', including two plenary talks, four contributed talks and the end-of-splinter discussion session.
Section~4 will summarise the splinter session as a whole.

\section{Block 1: `Multiplicity of cool stars -- Current Developments and Challenges'}

\subsection{Plenary Talk 1: `Detecting companions around cool stars'}

This plenary was presented and summarised here by \textbf{L\'ea Planquart}.
Companions around cool stars can be found at all mass ranges: while representing about $\sim20\%$ for red dwarfs in the lower mass range, this increases to $\sim 50\%$  for main sequence Sun-like stars and even higher for the most massive stars \citep{Offner2023}. The latter two, as they evolve through red giant branch phases (RGB, AGB, or RSG), imply a high number of cool evolved stars in multiple systems (binaries or higher multiplicity). To understand the impact of companions on the evolution of the components' properties (e.g., rotation, chemical composition), getting the orbital configuration is crucial.  

Observing 3D orbit motion from a fixed point of view implies projection, either on the plane of the sky, in the case of visual or astrometry binaries, or on the line of sight, in the case of spectroscopic or eclipsing binaries (EBs). Hence, to get the full set of orbital parameters and the individual masses, it is mandatory to use a multi-instrumental approach and combine different techniques. Below are listed some of the current main techniques and their caveat(s) when applied to cool-star primaries, whose dynamic and rich atmosphere induces one extra layer of complexity in extracting the Keplerian signature.

\subsubsection{Photometric binaries}

Characterizing orbital motion through photometry is often done through the transit method, where fitting the light curve gives access to the individual radii and effective temperatures of the components. Many large-scale photometric surveys exist; among them are OGLE -- which detected over 450,000 eclipsing or ellipsoidal binaries \citep{2016AcA....66..405S} --, \textit{Gaia} DR3 -- with about 530,000 candidates \citep{2023A&A...674A..16M} -- or \textit{TESS} for high-cadence photometry (hence short orbits or transits of a sub-stellar companion). One of the main challenges when dealing with a cool primary star is to disentangle the companion effect (eclipse) from intrinsic variability (e.g., periodic pulsation or dust ejections).  

One striking example of ambiguous eclipses is the family of long secondary period variables. These variables populating the D-sequence in the period-luminosity diagram  \citep{1999IAUS..191..151W} exhibit a long secondary period in their optical light curve together with a mid-infrared secondary eclipse \citep{2021ApJ...911L..22S}. Several scenarios exist to explain this periodic and chromatic behavior, involving a sub-stellar companion or non-radial pulsations \citep{2020MNRAS.492.1348T,2025A&A...703L..23D}, with no consensus reached yet.

\subsubsection{Spectroscopic binaries} 

Spectroscopic binaries are divided into two categories, SB2 and SB1. In an SB2 system, the radial velocity of each star can be computed, leading to a direct estimation of the mass ratio. In SB1, only the brightest component is seen; hence, only partial access to the mass through the mass function $f(m) = m_1^3\sin(i)/(m_1 +m_2)^2$, where $i$ is the unknown inclination angle between the orbital plane and the plane of the sky. 

 When dealing with cool stars, the main limitation is again to disentangle the companion effect (Keplerian orbit) from stellar variability, such as periodic pulsations and convective motions, that can lead to radial-velocity shifts of similar amplitude. Therefore, long-term (decade-long) radial-velocity monitoring is mandatory to filter out short-term variability, as well as an adapted model to fit and remove the effect of stellar pulsation (e.g., through Gaussian processes regression, as applied to extract the orbital parameters of Betelgeuse;  \citealt{2025ApJ...978...50M}).

\subsubsection{Astrometric binaries} 
  
As for spectroscopic detections, monitoring the astrometric motion of each component gives access to the mass ratio. While one component is too faint to be detectable, its presence still induces a periodic displacement of the system photocenter that allows one to characterize the orbit of invisible companions.  
Astrometric binary detection relies mainly on two space missions: ESA/\textit{Hipparcos} (1989--1993) and, more recently, ESA/\textit{Gaia} (2013--2025). Hence, one of the main limitations is the short time baseline covered by each mission. To remediate this, catalogues combining both missions exist (e.g. \citealt{2021AJ....162..186B} and \citealt{2022A&A...657A...7K}) but are restricted to primaries close and bright enough to have a reliable \textit{Hipparcos} astrometric solution.  

In that aspect, the next \textit{Gaia} release (GDR4) will be a game changer as it will produce an order-of-magnitude larger astrometric solution for multiple systems. The discovery of BH3, the $m_2 \approx 33 ~\rm M_\odot$  mass black hole around a cool primary \citep{2024A&A...686L...2G}, in pre-release illustrates the potential of \textit{Gaia} for detecting new families of unseen companions across the stellar mass range.

 \subsubsection{Interacting systems}

 During their evolution, stars in a binary (or multiple) system can interact and therefore affect their individual evolution. Systems caught in the act of interaction are important to study to understand mass transfer episodes. Interacting systems differ from quiet systems in that they bear, on top of their orbital motion, an additional observational footprint of mass transfer. Such an imprint can complicate the extraction of the orbital parameters.  

At sub-orbital scale, interactions with a companion lead to some observables, both on the donor and accreator side. On the donor/primary side, the gravitational attraction may induce a surface deformation, which gives rise to ellipsoidal variation in the light-curve or, for the closest systems, radius variation in interferometric observation. On the accreator side, mass transfer often occurs through the creation of an accretion disc, which gives rise to the following signatures: X-ray emission from its inner boundary layer \citep{2013A&A...559A...6L}, phase-resolved emission lines from the outer part and the bright spot (e.g., through the Doppler tomography method, \citealt{2001LNP...573....1M}), or flickering in high-cadence photometry.

The interaction type depends mainly on the orbital separation, $a$, and donor radius. Cataclysmic variables --made of a red dwarf and an accreting white dwarf-- have the shorter separation ($a \sim \rm 0.01\,au$), implying deformed primaries and mass transfer through Roche lobe overflow \citep{2003cvs..book.....W}. Symbiotic systems --made of a red giant and an accreting white dwarf-- have a wider range of separations ($a \in \rm 1\rightarrow\,  \sim100\,au$) to accommodate the larger dimension of the primary. For the wider system, the mass transfer is expected to be driven by wind Roche lobe overflow \citep{2007ASPC..372..397M}.  Another family of IB is post-AGB binaries, consisting of a post-AGB primary with a main-sequence companion with a jet-launching accretion disc and surrounded by a second-generation proto-planetary disc \citep{2025Galax..13...68V}.
Depending on the interaction mechanism, the evolution of the system can lead to a zoo of binary interaction products.

\subsection{Plenary Talk 2: `Observational clues of multiplicity-driven stellar evolutionary pathways'}

This plenary was presented by \textbf{Natalie Gosnell} and summarised by the session organisers.
The fate of binary systems largely depends on their orbital periods, though mass ratios and metallicities also play important roles. Close binaries (orbital periods $\lesssim$ 1000~d) may interact via mergers and mass transfers \citep{2008MNRAS.387.1416C}, while wider binaries may interact via wind Roche-lobe overflow \citep{2007ASPC..372..397M}. {\it Gaia} DR3 has given us a large sample of binaries to understand stellar interactions \citep{2024NewAR..9801694E}, with more systems expected from {\it Gaia} DR4.

Open clusters are great laboratories to understand the evolution of these IBs and compare them against single-star evolution. Using a color-magnitude diagram (CMD) of an open cluster, one can easily spot systems that stray away from single-star evolution. The most common of these anomalies are blue-straggler (BS) stars. Several studies have established that most of these systems form through mass transfer from a red giant or asymptotic giant branch star onto a main-sequence star \citep{2014ApJ...783L...8G,2015ApJ...814..163G,2019ApJ...885...45G}. This process leaves behind a binary consisting of a BS primary with a WD companion. The WD effective temperature and mass indicates how long ago the mass transfer occurred, and the evolutionary state of the giant star progenitor. BS stars have been found to rotate faster than other stars in the cluster due to mass transfer. 

This observation brings up the question of whether other stars with mass transfer show this characteristic. This is indeed true for another kind of system affected by mass transfer called Blue Lurkers \citep{2019ApJ...881...47L}. Blue Lurkers form through the same mass transfer pathways as BS stars, but the final mass transfer product is less massive than the main sequence turnoff, so the stars are “hidden” among the main sequence itself. These systems are not easy to identify but make up more than 10 percent of all spectroscopic binaries among the solar-type stars in M67 \citep{2019ApJ...881...47L}. Interestingly, one Blue Lurker has also been found to be a product of a merger in a triple system \citep{2025ApJ...979L...1L}.

Another class of anomalistic systems comprises the sub-subgiant stars. These systems lie below and away from the subgiant branch in an open cluster. These binary systems are magnetically active and fast rotators \citep{2022ApJ...925....5G}. In comparison, only 1 percent of single subgiant stars are rapid rotators \citep{2025AJ....169..309D}. 

The above examples provide us with simple observational clues of multiplicity-driven stellar evolutionary pathways. For stars with rapid rotation, close white dwarf companions, and heightened magnetic activity, we need additional checks to constrain whether these stars underwent interactions with another star(s). This understanding will further help to explore the branches of future evolution of these altered products.

\subsection{Plenary Talk 3: `Binary interactions with evolved giants and their progeny’}
\label{sec:plenary3}

This plenary was presented by \textbf{Onno Pols} and summarised by the session organisers.
The phenomenon of a companion interacting with an evolved giant star is relatively common, being experienced by about 30\% of all low-mass stars \citep{Mathieu2025}. The main pathways through which the interaction happens are through Roche-lobe overflow (RLOF, for periods shorter than $\sim 1000$ days) and through
stellar wind interaction (for wider orbits). Interaction can happen during the red giant branch (RGB) or the asymptotic giant branch (AGB). 

The progeny of these interacting systems are found throughout the H-R diagram, with post-RGB and post-AGB binaries representing systems with very recent interaction, sub-dwarf B stars being the outcome of systems that interacted at the tip of the RGB, and different groups of polluted stars (Ba stars, extrinsic S stars, CH and CEMP-s stars, and dwarf carbon stars) encompassing systems that interacted during the thermally pulsating AGB phase and that display the telltale enhancement of carbon and s-process elements.

Most of these classes of remnant systems share similar orbital properties, including orbital periods between 100 and $10^4$ days and substantial eccentricities. 
These orbital properties are also common in objects belonging to other classes, such as blue stragglers in old populations, S-type symbiotic binaries, and astrometric binaries with an M-dwarf star and a companion that is likely a white dwarf (WD) \citep{Shahaf2023}. 
The exception is C-rich dwarfs, which appear to show a bimodal distribution with an additional population at P < 5 days, similar to short-period WD + M-dwarf binaries \citep{Nebot2011}.

Despite being very common, the properties of these post-interaction systems are different from our expectation of how orbital evolution should proceed. First, the orbits of systems that experience RLOF are expected to be circularized due to tidal interaction. Additionally, RLOF from giant donors has often been assumed to be unstable in most cases, which should lead to common envelope evolution and the dramatic shrinking of the orbit. Finally, the orbits of systems that avoid RLOF and experience wind interaction should widen significantly. Hence, a gap is expected to appear in the period distribution around the period where RLOF starts to operate ($\sim 1000$~days) and orbits should become circular for $P<3000$~days \citep{Izzard2010,Nie2012,Abate2015}. As discussed above, observations directly contradict these two expectations.

Recent advances in our theoretical understanding indicate possible solutions to this mismatch. One commonly made assumption when treating RLOF is that red giants respond adiabatically to rapid mass loss and expand, causing unstable mass transfer. However, more realistic models show that the outer layers are superadiabatic and can thermally relax even at high mass-loss rates, which can make mass transfer stable \citep{Woods2011,Pavlovskii2015,Ge2015,Ge2020,Temmink2023}. In this case, the drastic orbital shrinkage of systems experiencing RLOF can be avoided.

In the case of wind interactions in systems with an AGB star donor, hydrodynamical simulations also reveal important features. A canonical assumption is that the outflow is isotropic, but that is often not the case for the slow, dense outflows of AGB stars. This gives rise to a mass-transfer mechanism named wind-RLOF, in which the slow, highly distorted outflow of the AGB star fills the Roche lobe \citep{2007ASPC..372..397M}. This produces larger accretion efficiency compared to Bondi-Hoyle accretion \citep{Mohamed2010, Chen2020}, enhanced angular momentum loss and potential orbital shrinking \citep{Jahanara2005,Chen2018,Saladino2018,Saladino2019}, and the possible formation of a circumbinary disc \citep{Chen2017,Chen2020}. Hence, the orbits of systems experiencing wind-RLOF might not widen as originally expected for wind interaction.

Finally, the assumption that tides always circularize orbits has also been subject to scrutiny. This expectation is based on the original equilibrium tide model with convective damping \citep{Zahn1977,Hut1981}, which is very well-tested for red-giant binaries \citep{Verbunt1995,Beck2018} and predicts very effective circularization of orbits at RLOF ($e<0.001$) except potentially for TP-AGB donors \citep{Phinney1992,Dewberry2025}. The conclusion is that an eccentricity-pumping process that counteracts tides appears to be required. The most prominent suggestions for such a mechanism consists of 1) mass loss and mass transfer varying along the orbit, such as enhanced wind mass loss at periastron \citep{Soker2000,Siess2014} or enhanced RLOF at periastron \citep{Vos2015,Hamers2019,Parkosidis2026}; 2) interaction with a circumbinary disk \citep{Waelkens1996,Dermine2013,Vos2015}; 3) dynamical interaction with a tertiary \citep{Perets2012,Toonen2020}; and potentially 4) WD birth recoils \citep{Izzard2010,El-Badry2018}.

Systems such as post-RGB/AGB binaries are important test systems \citep{Oomen2018,Moltzer2025} and display circumbinary discs \citep{VanWinckel2003,Kluska2022}.
The post-RGB binaries have periods consistent with stable RLOF \citep{Moltzer2025}. However, post-AGB orbits have lost additional angular momentum. Interactions with a circumbinary disc can reproduce the observed eccentricities provided that 1) the orbits are already fairly eccentric from the start, 2) circumbinary discs are much more massive than currently observed, and accretion from the circumbinary disc was inefficient \citep{2026Moltzer}. This suggests that post-AGB systems with massive circumstellar tori \citep{Khouri2022,Khouri2025} may be transitional objects to current post-AGB binaries and can potentially provide important constraints to understand orbital evolution.

In conclusion, while many open questions remain, significant advances in the last two decades are reshaping our understanding of the orbital evolution of IB systems and opening the way for their puzzling features to be theoretically explained.

\subsection{Summary Talks}

Along with the plenary talks selected by the splinter science organising committee, we selected 11 talks from submitted abstracts.
These talks were 5 minutes each, designed to introduce topics and prompt discussion for the later 'discussion session'.
7 were selected to present within the 'Current developments and challenges' section, covering a broad range of topics under the umbrella of stellar multiplicity with the remaining 4 to come in Block 2.

\textbf{Alexandra Boone} was first to present, discussing the fundamental parameters of low-mass doubly eclipsing SB2 binaries. Alexandra uses the false positives from the 'Searching for GEMS' (Giant exoplanets around M-dwarf stars; \citealt{2024KanodiaGEMS}) survey, characterising EBs found in a 200 pc volume-limited sample combining \textit{TESS} and ground-based photometry and spectroscopic data from the Habitable Zone Planet Finder. The survey is still collecting data but currently has 94 targets of which 45 are double-lined EBs which is aimed to help provide a more robust comparison of the mass-radius relationship to theoretical models, aiding the constraining of the effect of radius inflation. The preliminary analysis of the double-lined binary TOI-5658 was presented, displaying the RV and light curve analyses and the derived parameters of the system including mass and radius ratios.

\textbf{Florian Driessen} introduced the topic of mass-loss in AGB binary systems, presenting results from theoretical hydrodynamical simulations. AGB stars and their pulsations and mass loss were introduced, with the existence of some of these systems as binaries raising the question of the effect of multiplicity upon the mass loss. Although a presented sample of AGB stars seems to fit an empirical mass-loss relation derived from CO lines \citep{2010DeBeck}, 2-D hydrodynamical simulations of a pulsation-dust-driven AGB wind + M-dwarf/white dwarf system showed that extreme interaction can pump mass-loss from 10 to 100 times greater. With most of the previously presented sample being found to be multiple star systems this suggests that low mass-loss rates for a given pulsation period could be minimum values due to being a single AGB star, whereas maximum values could indicate multiplicity pumping the mass-loss to higher rates. Mass-loss being pumped by multiplicity would thus be an important effect to consider not just for empirical calculations of mass-loss but for models of stellar evolution and nucleosynthesis as well.

We then heard from \textbf{Mark Giovinazzi}, presenting a volume-limited sample of companions to accelerating stars from \textit{Hipparcos} and \textit{Gaia}. His work focuses on systems who are cold, wide-orbit binaries, an as of yet relatively sparse population of companions. He discussed how combining radial velocity and astrometry breaks the degeneracy between mass and inclination, gaining dynamical masses and refined orbits. On the astrometric side exploiting the change in proper motion observed between the \textit{Hipparcos} mission (epoch of 1991.25) and \textit{Gaia} (epoch of 2016.0), an astrometric acceleration caused by the secondary star on its primary can be obtained leading to a direct inference on the  secondary mass. By using the 20 pc volume sample observed by both missions (\textit{Hipparcos}-\textit{Gaia} Catalogue of Accelerators (HGCA); \citealt{2018Brandt,2021Brandt}), they found 206 accelerating stars, calculating new dynamical masses for 50+ binaries and newly characterising 2 brown dwarfs. With 24 accelerating stars with no previous RV baseline now being monitored and \textit{Gaia} DR4 forecast to increase the sensitivity of HGCA, Mark and his team hope to increase these numbers further and detect new orbits, trends and occurrences within their sample. 

\textbf{Michael Greklek-McKeon} then took us to the planetary, presenting confirmation of three Earth-sized planets in the most compact known planet-hosting M-dwarf binary. He presented the TOI-2267 system, a M-dwarf-M-dwarf binary with an 8\,au separation, which had 3 \textit{TESS} candidate planets. Accounting for the dilution effect of both binary hosts, his team used ground-based photometry to verify the planets, finding orbital periods of 2.0, 2.3 and 3.5 days \citep{2025Zuniga,2026Greklek}. Two orbital configurations were presented for the system. The first involving both stars having orbiting planets would make TOI-2267 the closest planet-hosting binary system and challenge planet formation theory to account for formation around what would be two truncated disks. Option two would involve one star of the binary hosting all three planets which would require the 'most compact' exoplanetary system. They are working on finding which configuration is correct through transit chromaticity, TTVs or a fully resolved transit observation.

\textbf{Axel Hahlin} looked at M-dwarfs as orbiters rather than hosts, presenting characterisation of EB benchmark stars and their M-dwarf companions. As part of the \textit{PLATO} WP125500 - Benchmark stars working programme, his team seeks to characterise Eclipsing Binary with Low-Mass companion (EBLM) systems, increasing the number of model-independent benchmark stars for the mission, useful for performing end-to-end verification of stellar pipelines. They characterise the EBLMs using spectroscopic and photometric observations, obtaining the dynamics of the orbiting M-dwarf by constructing a mean-spectra (from which chemical composition can also be measured), removing the primary signal and using cross-correlation. They use the programme TEB \citep{Miller++20mn} to fit fluxes and flux ratios to obtain the effective temperatures of each star. Overall, they reach values of mass, radius and temperature with a precision of < 1 \% and results for their first published target can be found in \cite{2026Adshead}.

\textbf{Fabian Kaczmarek} provided further observational insights of EBs both active and non-active, using interferometry. His group has worked on symbiotic binaries, those with a white dwarf component and a cool red-giant donor, whose large size and brightness in the NIR makes them ideal targets for optical interferometric study. Using this technique, they can find the angular diameter of the donor, its size and through knowing its Roche lobe radius determine which mass-transfer process dominates. However the main subject of the presentation was novae, a switch in subject caused by the explosion of some of the symbiotic stars the team were observing. They observed the expansion, geometry and kinematics of the post-outburst envelope of RS Oph and are tracking the change in diameter of the donor star in a pre-outburst symbiotic star T CrB \citep{2026Norris}, which is expected to go Nova.

Then with the last contributed talk of the session, \textbf{Jaroslav Merc} continued the theme of interactions, probing them in cool stars beyond the Roche Lobe overflow. Giant-branch stars in IBs along with regular Roche lobe overflow also interact through other mechanisms such as through wind, with a prime example being the previously introduced symbiotic binaries \citep{2025Merc}. At orbital periods of hundreds of days, these wide binaries (e.g. R Agr and Mira) can show evidence of mass transfer from both wind and Roche Lobe overflow. Therefore, his team performed interferometric analysis of 13 symbiotics with VLTI/PIONIER \citep{2025MercBoffin,2025Boffin} and mostly found them to not be filling their Roche Lobes. This raises the question of what is causing the observed ellipsoidal variability and if modifications must be made to their Roche potentials.

Finally, the block closed with a small 'poster-pop' session where 9 presenters were given thirty seconds to highlight their work and the posters they were presenting at the conference.
We were presented with a variety of results spanning observations and theory.
The complex architecture of a multiple brown-dwarf system was spotlighted, through the $\sim$ 45 Myr system 2M1510.
The system NGC 1333 in an intermediate-density star-forming region was examined using multi-filter HST imaging to identify substellar companions to more than a dozen stars within.
We heard of efforts using VHS and DES to find T-dwarf candidates around \textit{Gaia}-detected primary stars, with a final sample of 10 binary pairs of which 4 were unreported.
\textit{Gaia} was a key theme of our poster-session with multiple projects using it and looking to the future DR4. One presented their work where simulated \textit{Gaia} data was combined with cutting-edge ground-based astrometric observations from the NTT using the open-source Python joint-parameter solver \texttt{plxmaxpy} in a technique they are hoping to leverage with the upcoming DR4 to ensure well-calibrated precise results.
Another used \textit{Gaia} pre-selection to perform a multiplicity survey of nearly 600 stars with Gemini and Keck, using it to explore the Galactic field, the history of OB associations and how environment effected their formation.
Also looking ahead to DR4, a pilot astrometric survey using RECONS, spectroscopy and MAROON-X to obtain 54 stellar orbits and detect 100 stellar companions, with hopes to provide complete orbits and companions for every M-dwarf when combined with \textit{Gaia}’s upcoming release.
The young spectroscopic T~Tauri binary DQ Tau, which undergoes periodic accretion bursts, was studied using observations from UVES and X-SHOOTER at the Very Large Telescope to study the characteristics of its outflow including the first-time wind launch radius has been directly observed to be changing.
DQ Tau also came under examination using archival data to investigate this pulsing accretor. Multiple decades worth of RVs and transits were leveraged to provide statistical inferences not capable within a single season of observation, with results such as supporting inner-disk variability and detecting a low mass circumbinary companion.
Finally, we saw triple stars examined, with the presentation of different scenarios for the formation of compact hierarchical triples, exploring dynamical effects such as disk instability, core fragmentation, cloud collisions and dynamical interactions.


\section{Block 2: 'Preparing for the future -- Future missions and Cool Stars Multiplicity'}

\subsection{Plenary Talk 4: `Binary Stars and RR Lyrae Variables with the Vera Rubin Observatory's LSST: Preparing for a New Era of Cool Star Science'}

This plenary was presented by \textbf{Kelly Hambleton} and summarised by the session organisers.
When discussing the upcoming tranche of new and future missions, one with much potential for Cool Star science is the Vera C. Rubin Observatory, formerly the Large Synoptic Survey Telescope (LSST), an 8.4 m telescope in Cerro Pach\'on, Chile \citep{2019Ivezic}.
Having its first light on 3rd June 2025 and commencing full survey operations in July 2026, Rubin LSST will be one of the foremost resources for our community over the next 10 years through its Wide-Fast-Deep (WFD) 10-year survey.
The survey will be performed with LSSTCAM, a 3.2 gigapixel camera with a 1.55\,m lens and a 9.6 square degree field of view, wide enough to image seven moons across the face of the camera and with a very high resolution.
LSSTCAM will be observing in six photometric bands from the near-UV to near-IR, enabling a wide variety of science goals. 
Covering 18,000 square degrees of sky the WFD survey will perform around 837 visits per pointing over 10 years, with Deep Drilling Fields (DDFs) having an order of magnitude more visits. 
A single visit will be able to go to a depth of $r$ magnitude of $\sim$ 24.7\,mag and when co-added, a depth of $\sim$ 27.5\,mag.

Data from Rubin LSST has already been released, with first-look images and DP1 commissioning data (using the commissioning camera, which uses only 9 CCDs, i.e. 1 rack) already available for limited fields \citep{2026Vera}.
DP2 science validation data will be released September 2026 \citep{VeraDP2}, with the first full data release DR1 scheduled for June 2028.
A description of the pipeline products available for Rubin LSST can be found in \cite{VeraPipe}.
We expect Rubin LSST to have up to 10 million alerts per night, taking in 10--20 TB of raw data per night, $\sim$ 60 PB over 10 years.

Within the Rubin LSST team, the science cases for cool multiple stars falls under the Transient and Variable Star (TVS) Science Collaboration, including IBs, EBs and RR Lyrae stars \citep{Hambleton_2023}.
Rubin LSST will seek to identify new novae, dwarf novae and nova-likes as well as observe recurrent novae.
It will seek to generate a census of IBs and identify magnetic cataclysmic variables \citep{2025Buckley}.
Generally, it should be capable of extended observations of IBs in all states and bridge the gap between the currently known $10^4$ IBs and $10^7$ EBs, as well as be able to identify those that are eclipsing.

On non-interacting EBs, Rubin LSST will also be a great resource.
Through deriving flux-calibrated colours it will be able to derive absolute temperatures and using luminosity calibrate trigonometric parallaxes \citep{Hambleton_2023}.
Using \textit{Gaia} distances this will result in absolute parameters, including masses for contact binaries.
Late-type contact binaries will also be able to have their distance derived using period-luminosity relations.
Rubin LSST will be able to achieve a near-complete census of short-period eclipsing, ellipsoidal and contact binaries, and potentially detect coalescence events.
Through simulating for Rubin, using LSST pointings on 2600 \textit{TESS} binary light curves, Rubin LSST should recover 65\,\% of EBs with this being a near-complete census for periods less than 1 day \citep{2023Prsa}.
In total, Rubin LSST should observe between 10$^6$ and 10$^7$ EBs.
The initial DP1 has already resulted in papers on EBs with \cite{2025choi} identifying two EBs in 47 Tucanane.
Moreover, using DP1 data, \cite{2025cordoni} identifying unresolved binaries with a mass ratio more than 0.7 in 47 Tucanae, providing additional evidence that there are more binaries in the outer regions of the cluster, potentially due to disruption in the cluster core.

Rubin LSST observations of RR Lyrae stars will allow the exploration of intriguing science, tracing galactic structure, stellar populations and chemical evolution \citep{Hambleton_2023}.
Its multiband photometry will enable the use of Period-Wesenheit relations to cancel extinction and provide photometric metallicity estimates from colour-period metallicity relations. 
The additional color information will simplify classification into RRab and RRc subclasses and enable us to distinguish between RR Lyrae and contact binaries.
Up to 100,000 RR Lyrae stars are anticipated to be observed using Rubin LSST and similar simulations to \cite{2023Prsa} using \textit{TESS} light curves are being performed.
Early Rubin LSST science is also achieving results with \cite{2026Ngeow} identifying 600 RR Lyrae stars and constraining their metallicities using DP1 data, Gaia periods and templates.

Although the full Rubin LSST sample will take 10 years to survey, deep drilling fields (DDF) will provide a yearly output of data \citep{Hambleton_2023}.
In each DDF, 2000 data points will be generated each year, with the COSMOS DDF obtaining an average of 4800 points a year.
The field locations have been selected to overlap with the \textit{Roman} and \textit{Euclid} missions which will help improve cadence, crowded field identifications and lower extinction sensitivity.
Within these DDFs, 10$^4$ RR Lyrae stars and 10$^5$ EBs are expected.

\subsection{Plenary Talk 5: `Multiple stars with \textit{PLATO}'}

This plenary was presented by and summarised here by \textbf{John Southworth}.

\subsubsection{Multiplicity in cool stars}

Multiplicity is an unavoidable feature of stars -- the majority of stars have at least one companion -- and also a vital tool for measuring their physical properties. Multiplicity is a natural outcome of star formation; the observed binary (and multiplicity) fraction, and the mass, mass ratio and orbital period distributions, provide constraints on the different star formation mechanisms \citep{Offner2023}. Binary stars are common so all stellar populations have binaries, whether we like it or not, which must be accounted for in studies of these populations. Binary evolution, especially mass transfer,  creates some of the most interesting objects in the universe, including common-envelope phases and planetary nebulae, cataclysmic variables, novae, supernovae, X-ray binaries, gamma-ray bursts, and the progenitors of gravitational-wave events.

There are many methods to detect and quantify stellar multiplicity. These include finding eclipses in the light curves of stars \citep[for example Algol;][]{Goodricke1783}, astrometry \citep[for example Sirius;][]{Bessel1844mn}, spectroscopy (where variations in radial velocity with time indicate an orbiting companion, or a single spectrum may show two or more sets of spectral lines), and direct imaging \citep{Herschel1782rspt}. These methods can find both binary stars and higher-order multiples, sometimes in several ways. As an example, the extremely bright star $\beta$ Aurigae is an eclipsing, spectroscopic \emph{and} astrometric binary \citep{Me++07aa}. Systems containing three or more stars may show dynamical effects such as apsidal motion and the disappearance of eclipses \citep{Glowacki+24aca}.

It is now well-established that multiplicity depends on stellar mass \citep{DucheneKraus2013,Offner2023}. The multiplicity of O-stars is high: \citet{Sana+14apjs} found that $0.91 \pm 0.03$ of a sample of O-stars had companions. At the other end of the main sequence, \citet{Shan++15apj} used EBs observed by the \emph{Kepler} missions to deduce that M-dwarfs have on average $0.11 \pm 0.03$ companions with orbital periods between 1 and 90\,d. The binary and multiplicity fractions decrease with age due to disruption by encounters with nearby stars \citep{Jaehnig+17apj}, and increase to lower metallicity \citep{Badenes+18apj} due to opacity affecting radiation losses during star formation.

\begin{figure}[t!]
	\centering
	\includegraphics[width=\columnwidth]{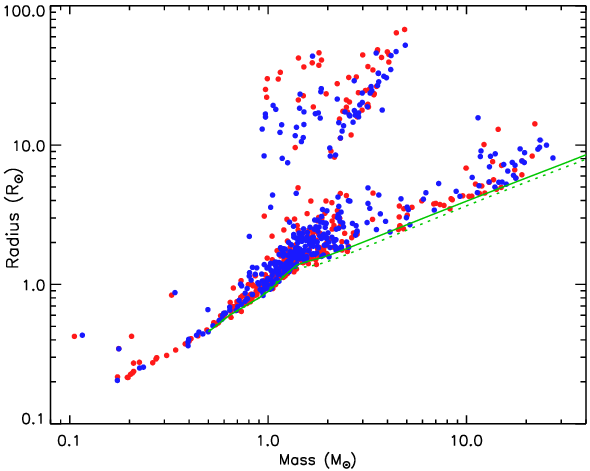}
	\caption{\label{fig:debcat} Mass-radius plot for all stars in the Detached Eclipsing Binary Catalogue 
      \citep{Me15aspc} as of 29th July 2026. Blue points are the primary (i.e. brighter and/or more massive) companions of the binary, red points are the secondary stars, and 
      the green lines show the ZAMS for solar (unbroken line) and half-solar (dashed line) metallicity.}
\end{figure}

EBs are arguably the most interesting binary systems because from their light and radial velocity curves it is possible to measure their masses and radii directly and to high precision. Detached systems, i.e.\ those whose stars are sufficiently far apart to have experienced no mass loss or transfer, are particularly valuable as they have evolved like single stars so can be compared to theoretical models of single-star evolution. The recent availability of \emph{Gaia} parallaxes means we can obtain their luminosities and thus effective temperatures directly as well \citep{Miller++20mn}. The radii of cool stars in EBs are consistently observed to be larger than predicted by up to 15\%: this is called the radius discrepancy, has persisted for over 30 years \citep{Hoxie73aa}, and remains unsolved \citep{Swayne+24mn}. The mass-radius diagram of detached EBs with precise mass and radius measurements in shown in Fig.\,\ref{fig:debcat}.

\subsubsection{Multiple stars with the \textit{PLATO} satellite}

The \textit{PLATO} (PLAnetary Transits and Oscillations of stars) mission is a space-based telescope scheduled for launch by the European Space Agency (ESA) in January 2027. It will travel to the second Lagrangian point in the Sun-Earth system and obtain light curves of nearly 400,000 stars for multiple years. Its aim is to find extrasolar planets by the transit method, with a particular emphasis on Earth-like planets (rocky planets in the habitable zones of FGK dwarfs). It will observe the LOPS2 field \citep{Nascimbeni+25aa}, which covers 2137\,deg$^2$ in the southern hemisphere, for a minimum of 2 years, after which it may stay on this field or move to a northern one. \textit{PLATO} contains 26 small cameras (12\,cm pupil) of which 24 are dedicated to science and the remaining two are fast cameras used for maintaining pointing plus obtaining light curves in two colours. The distance from Earth to \textit{PLATO} means telemetry is limited and thus only pre-selected targets will be observed.

Of specific interest to this conference is the `P4 sample' to be observed by \textit{PLATO}, which must contain at least 5000 M dwarfs with brightnesses $V < 16$\,mag \citep{Prisinzano+26aa}. Low-mass stars are important for transit surveys because their small sizes mean smaller planets can be detected, and their low luminosities mean the habitable zone is much closer to the star (hence at shorter orbital periods). The P4 sample will contain several tens of EBs, most not currently known. More EBs have also been selected for observation via the scvPIC mechanism \citep{Zwintz+26exa}.

John Southworth has formed the \textit{PLATO} Multiple Star Working Group (MSWG\footnote{\url{https://www.astro.keele.ac.uk/jkt/plato/index.html}}) as an implementation of Work Package 161,000 in the \textit{PLATO} mission. Its aim is to co-ordinate all work on binary and multiple stars within the \textit{PLATO} consortium, including obtaining \textit{PLATO} data, acquiring ground-based follow-up observations, working with the transiting planet identification and characterisation groups within \textit{PLATO}, and organising the work involved. The MSWG has over 100 members worldwide and anyone can join by emailing John Southworth with their name and a brief description of what science they are interested in. It is expected that there will be excellent data for far more binary stars than the MSWG can analyse themselves, so new members are welcome! The MSWG is currently compiling a White Paper covering all the science cases they intend to pursue with \textit{PLATO} data, which will be published in the \textit{PLATO} Special Issue in the journal Experimental Astronomy.

The \textit{PLATO} mission includes a Complementary Science component through which applications can be made to observe any object in the LOPS2 field via the Guest Observer (GO) mechanism run by ESA. The deadline for the first round of GO applications was 21st May 2026 and the outcome of the applications is expected by the end of August 2026. As a result, the MSWG will have access to light curves of binary and multiple star systems via three routes: the \textit{PLATO} planet search survey; the scvPIC mechanism; and the successful GO applications submitted by the group. The work at Keele University will focus on cataloguing all EBs in the LOPS2 field (Kutluay et al., in prep), analysing them using automated methods \citep{OverallMe25rasti} and organising detailed studies of the most interesting objects.

\subsection{Summary Talks}

After the last of the splinter plenary talks it was time for our contributed five-minute talks, of which 4 were scheduled for the `Preparing for the future' block.
First, we had \textbf{Zachary Hartman} who presented the updating of the MUGSHOTS survey. Standing for MUltiplicity of Galactic Stars in the Halo and Orbit Tracking Survey, MUGSHOTS aims to observe the multiplicity of the galactic halo and has observed 700 stars since 2014 of which 10.2\,\% are in multiples. This is much lower than previous estimations which would have impacts on fields from binary and star formation to dark matter content. They look to the future and to \textit{Roman}, inside whose field \textit{Gaia} denotes 415,207 thick disc stars and 114,381 halo stars, a number which would imply 10,100 and 1503 binaries respectively. In five years \textit{Roman} could allow the analysis of 500$\times$ the number of stars as the previous survey, with the team of what will be MUGSHOTS+ predicting this to be an underestimation, and through its sensitivity will enable the discovery of fainter M-dwarf halo stars from the current limit of a {\rm $M_G$} of $\sim$ 15\,mag down to $\sim$ 20\,mag.

Next, we explored the magnetism of interacting systems in open clusters with \textbf{Silva J\"arvinen}. She introduced us to blue and yellow straggler stars (BSS \& YSS; and although not focused on, red stragglers), stars that defy the usual evolution with age within a cluster. These systems are derived from stellar interactions, and 3D MHD simulations propose such systems may exhibit strong magnetic fields and rapid rotation following a merger \citep{2019Schneider}. Using the \textit{Gaia} DR2 catalouge of BSSs \citep{2021Rain}, 5 BSS and 3 YSS stars were identified in open clusters of very different ages and metallicities finding no difference in magnetic field strength and cluster characteristics \citep{2025Hubrig}. They find that detected magnetic field strengths in YSS stars are comparable in strength to BSS stars indicating that magnetic fields created by the interaction process remain stable across evolutionary timescales. They now seek to further increase reliability of these inferences and determine properties through spectropolarimetric observations, a HARPS survey of 30 BSS/YSSs and possible observations of 4 BSSs and 4 YSSs in the \textit{PLATO} LOPS2 field.

\textbf{Sara Mu\~noz Torres} presented \textit{Euclid}'s first resolved substellar system, a rare wide L/T binary in the field. With a binary fraction between 5-25 \% and a tendency for separations less than 10\,au these objects are not very common. \textit{Euclid}'s deep field survey with a high spatial resolution of 0.1"/pixel in the VIS instrument provides a great opportunity to look for substellar companions around young, cool dwarfs. They present the newly discovered E271934, spatially resolved by \textit{Euclid}. Photometrically its magnitudes in the $Y$ and $J$ bands are expected of L-T transition objects but find an unusually wide separation of 70\,au, its 0.532 arcsec angular separation resolved thanks to \textit{Euclid}'s excellent spatial resolution. Using spectra from \textit{Euclid} and Gran Telescopio Canarias, along with synthetic spectra from SpeX Standards the two stars were determined to be L4--L5 and L9--T0 respectively, providing an excellent target for future follow-up and demonstrating the ability of \textit{Euclid} to resolve and characterise substellar binaries.

Finally, \textbf{Thomas Vandal} closed out the presentations by talking about using \textit{JWST} to constrain the multiplicity of the coldest brown dwarfs between 0.5 and 1000\,au. Focusing on Y-type brown dwarfs, he highlighted that multiplicity seems to decrease with lower mass \citep{2018Fontanive,2023Fontanive}, but with only a few Y dwarfs observable by \textit{Hubble}, asked if this was due to HST's low angular separation. They proposed using \textit{JWST} to probe the short separations (< 100 mas) of 22 Y-dwarfs for cold companions to constrain this binary frequency, expecting 1-3 detected companions. They first detect the first Y+Y-dwarf binary (WISE-0336) via ePSF modelling with a separation of only 1\,au. Then using Kernel Phase interferometry (using the telescope itself as a result of many, small telescopes interfering with each other) to unlock higher angular resolution they went through the rest of the survey, finding no further companions and constraining the binary fraction of Y-dwarfs to 5\,\%. They hope to expand these efforts through a wide companion search and through further \textit{JWST} observations, to constrain the predicted binarity rate further.

\subsection{Discussion Session}
The splinter hosted a discussion session where we took questions from the community and discussed the current status and future goals for the science of multiplicity of cool stars. We summarise the session below.

\subsubsection{In what configurations do cool stars in binaries behave as single stars?}

One of the most discussed issues in stellar evolution is when to consider non-interacting binary stars as a true proxy for a single star's evolution. It becomes important when we try to build mass-radius relations using EBs or when we calibrate asteroseismic scaling relations using red giant binaries. The answer to this question varies greatly across different subfields and is dependent on opinions on a certain phenomenon in stellar evolution. It primarily depends on the tidal deformation of the stars, which itself depends on the stellar masses in the binary, as well as the period and eccentricity of the binary. On crude scales and for solar-type main-sequence stars, circular systems with periods on the order of a few tens of days can be treated as single stars. But it is still an open question to explore.

\subsubsection{How complete are current multiplicity statistics of cool stars?}

The completeness of the multiplicity statistics is limited by instrumentation. The Pervasive Overview of ``Kompanions'' of Every M dwarf in Our Neighborhood (POKEMON) survey (\citealt{2026AJ....172...40C}; and references therein) has been looking at M dwarfs, and with every new observation they are discovering more and more companions. Even if we push the boundaries for completeness, we need to complete our understanding of a strict boundary between planets and the lowest-mass stellar objects. Until then, completeness is elusive. 

\subsubsection{What instrumentation is most needed?}

The need for a survey instrument that can provide high-resolution spectroscopy with long baselines was highlighted during the session. Though not high-resolution, the upcoming SDSS-V and {\it Gaia} DR4 are something that the community looks forward to.

\subsubsection{Should we as a community need standardization of datasets, catalogues, and codes for analyzing multi-star systems?}
In this era of big data, multiplicity studies demand combining different observational studies to improve our theoretical models. A major issue that arises during this process is inconsistency in the reporting of the different observational properties and parameters. Further, there is a need to discuss or set sensitivity limits to compare different surveys.  At the same time, there have been cases where data formats and observed parameters have been homogenised. For example, DEBCat \citep{Me15aspc} has compiled precise masses and radii for hundreds of individual systems reported in the literature. Similarly, the Multiple Star Catalogue (MSC), Washington Double Star Catalogue (WDC), and the Ninth Catalogue of Spectroscopic Binary Orbits (SB9) have compiled a large number of multiple-star systems over the years. But the issue of consistency persists because of the complexity of binary and multiplicity analysis. The same reasoning stops us from having a single self-sufficient code for binary or multiplicity analysis. But the availability of different codes ensures that we can test the accuracy of our estimates using different methodologies. Even with all the above issues, the community can benefit from a single database where we can provide access to all the catalogues and codes. We, therefore, created a Multiplicity Database\footnote{\url{https://stellarmultiplicity.github.io/database/}} that hosts links to various catalogues and codes related to binary or multiple stars. The website is community-driven in the way that everyone can suggest catalogues and codes using Google Forms so that the larger community can benefit.

\section{Conclusions}

With the renaissance of stellar multiples in full flow, their study is sure to continue to be invaluable for the field of low-mass stars.
We thoroughly examined the types of stellar multiple and their identification and characterisation.
The dynamic present of the  field was discussed, detailing advances in topics as wide-ranging as binary interactions, stellar evolution and the exploration of the substellar.
We then looked to the future, with a focus on Rubin LSST and \textit{PLATO} as well as \textit{Roman}, \textit{Gaia}, \textit{Euclid},\textit{JWST} and a wide variety of cutting-edge surveys and techniques that promise intriguing future results.
Finally, the splinter concluded by a lively discussion of the community, discussing the behaviours of cool objects in binaries, the completeness of multiplicity statistics, needed instrumentation and the standardization of datasets, catalogues and codes, resulting in the creation of a Multiplicity Database.
Whether through created connections, inspiration or generated resources, we look forward to seeing the impact of our splinter as we go into the bright future of the Multiplicity of Cool Stars and their Evolution.

\section*{Acknowledgments}
{We thank the organisers of Cool Stars 23 for the selection of this splinter and in all their help in preparing and running the session.
MIS is grateful for funding from STFC via Grant Number ST/X000885/2 and from the GB Sasakawa Foundation via grant number 6955. JS and AM acknowledge support from STFC under grant number ST/Y002563/1.
DMS has received funding from the European Union (ERG grant no. 101040160 ``LSP-MIST''). Views and opinions expressed are however those of the author only and do not necessarily reflect those of the European Union or the ERC. Neither the European Union nor the granting authority can be held responsible for them.}

\bibliographystyle{cs23proc}
\bibliography{example.bib}

\end{document}